\documentclass[hyper]{prop2015}
\usepackage[english]{babel}

\usepackage{amsmath,amssymb,amsthm}
\usepackage{amsfonts}
\usepackage{mathrsfs}
\usepackage{bbm}
\usepackage[all,knot,cmtip]{xy}

\usepackage{tikz}
\usetikzlibrary{arrows}
\usetikzlibrary{patterns}
\usetikzlibrary{decorations.markings}

\newcommand{\der}[1]{\frac{\partial}{\partial #1}}

\newcommand{\eand}{{~~~\mbox{and}~~~}}
\newcommand{\myxymatrix}[1]{\vcenter{\vbox{\xymatrix{#1}}}}

\newcommand{\di}{\mathrm{i}}
\newcommand{\dd}{\mathrm{d}}
\def\slasha#1{\setbox0=\hbox{$#1$}#1\hskip-\wd0\hbox to\wd0{\hss\sl/\/\hss}}
\newcommand{\id}{\mathrm{id}}

\newcommand{\acton}{\vartriangleright}

\newcommand{\frg}{\mathfrak{g}}

\newcommand{\FR}{\mathbbm{R}}     

\newcommand{\eps}{{\varepsilon}}			

\newcommand{\langlec}{\prec\hspace{-0.5mm}}
\newcommand{\ranglec}{\hspace{-0.5mm}\succ}

\newcommand{\epsb}{{\bar{\varepsilon}}}			

\newcommand{\vol}{{\rm vol}}

\newcommand{\sLie}{\mathsf{Lie}}
\newcommand{\sG}{\mathsf{G}}

\newcommand{\NN}{\mathbbm{N}}     			

\newcommand{\CF}{\mathcal{F}}
\newcommand{\CH}{\mathcal{H}}
\newcommand{\CG}{\mathcal{G}}
\newcommand{\CI}{\mathcal{I}}

\newcommand{\dpar}{\partial}     			
\newcommand{\dder}[1]{\frac{\dd}{\dd #1}}   		

\newcommand{\au}{\mathfrak{u}}
\newcommand{\asu}{\mathfrak{su}}
\newcommand{\aso}{\mathfrak{so}}
\newcommand{\aspin}{\mathfrak{spin}}
\newcommand{\astring}{\mathfrak{string}}
\newcommand{\CN}{\mathcal{N}}
\newcommand{\CL}{\mathcal{L}}
\newcommand{\sL}{\mathsf{L}}
\newcommand{\sEnd}{\mathsf{End}\,}
\newcommand{\RZ}{\mathbbm{Z}}     			
\newcommand{\sSpin}{\mathsf{Spin}}
\newcommand{\sString}{\mathsf{String}}
\newcommand{\sU}{\mathsf{U}}
\newcommand{\sSU}{\mathsf{SU}}

\category{Proceedings}
\title[Higher Structures, Self-Dual Strings and 6d Superconformal Field Theories]{Higher Structures, Self-Dual Strings\\ and 6d Superconformal Field Theories}
\subtitle{\href{http://www.maths.dur.ac.uk/lms/109/index.html}{LMS/EPSRC Durham Symposium on Higher Structures in M-Theory}}
\author[Christian S\"amann]{Christian S\"amann\inst{a}\footnote{Corresponding author e-mail:~\href{mailto:c.saemann@hw.ac.uk}{\textsf{c.saemann@hw.ac.uk}}}}
\address[1]{Maxwell Institute for Mathematical Sciences and Department of Mathematics, Heriot--Watt University, Edinburgh EH14 4AS, United Kingdom; EMPG--19--07}

\begin{acknowledgements}
I would like to thank Lennart Schmidt for a pleasant collaboration on the papers~\cite{Saemann:2017rjm,Saemann:2017zpd}, on which this review is mostly based. I am also grateful to all the participants of the \href{http://www.maths.dur.ac.uk/lms/109/index.html}{LMS/EPSRC Durham Symposium on Higher Structures in M-Theory} and of the workshop \href{https://ims.nus.edu.sg/events/2018/wstring/index.php}{String and M-Theory: the New Geometry of the 21st Century} for contributing to two inspiring workshops and many useful discussions. 
\end{acknowledgements}

\shortauthors{C.~S{\"a}mann}

\begin{abstract}
I summarize and discuss some recent results on formulating actions of six-dimensional superconformal field theories using the language of higher gauge theory. The latter guarantees mathematical consistency of our constructions and we review crucial aspects of this framework, such as $L_\infty$-algebras and corresponding kinematical data given by higher connections. We then show that there is a mathematically consistent non-Abelian extension of the self-dual string equation which satisfies many physical expectations. Our construction favors a particular higher gauge group leading us to higher principal bundles known as string structures. Using these, we manage to formulate a six-dimensional action which shares many properties with the famous $(2,0)$-theory but also still differs from it in some key points.
\end{abstract}
\shortabstract
\begin{document}
\maketitle

\section{Introduction}

Among the classical and quantum field theories, conformal field theories are particularly interesting and important. The most prominent example is perhaps maximally supersymmetric Yang--Mills theory in four dimensions. This theory has also been dubbed ``the harmonic oscillator of the 21st century.'' It is a very useful toy model for quantum computations in QCD due to its large amount of symmetries, which heavily constrain the theory's properties, simplifying the computations.

A conformal field theory on $\FR^{p,q}$ is invariant under the Lie algebra $\aso(p+1,q+1)$ of conformal symmetries. If we are now interested in supersymmetric conformal field theories (which we are again for their simplicity), we need to extend this Lie algebra to a super Lie algebra. Since the work of Nahm~\cite{Nahm:1977tg}, we know that such supersymmetric extensions are only possible for $p+q\leq 6$. Examples for $p+q\leq 4$ have been known for some time and the above mentioned 4$d$ super Yang--Mills theory is an example for $p=3$, $q=1$. It was commonly thought that $p+q=4$ was the maximum for interesting (i.e.~interacting) quantum field theories.

In his contribution to the conference ``Strings 95,'' however, Witten~\cite{Witten:1995zh} used string theory considerations to argue that there is an interesting six-dimensional superconformal field theory which is by now often referred to as the ``(2,0)-theory.'' This theory arises when compactifying type IIB string theory on $\FR^{1,5}\times $K$3$. The vacuum moduli space of this string compactification contains singularities at which 2-cycles in the K3 collapse to points. Wrapping a D3-brane about these 2-cycles leads to a string in $\FR^{1,5}$, which produces a self-dual 3-form $H=\dd B=*H$ as argued in~\cite{Witten:1995zh}. At the singularities of the moduli space, the string becomes massless and supergravity decouples. We are left with an interacting six-dimensional superconformal field theory with $\CN=(2,0)$ supersymmetries. 

This (2,0)-theory also arises on the world-volume of parallel M5-branes~\cite{Strominger:1995ac,Dasgupta:1995zm,Witten:1995em}, where the self-dual strings are the one-dimensional boundaries of M2-branes ending on the M5-branes. The importance of a good understanding of this theory becomes immediately clear when noting that it restricts to the above mentioned super Yang--Mills theory after a compactification down to four dimensions. In fact, Witten called the (2,0)-theory ``the pinnacle'' of the web of string dualities, since it determines many dualities in four and lower dimensions.

It is a widely held belief that contrary to most commonly studied quantum field theories, the (2,0)-theory does not allow for a classical description in terms of a Lagrangian except in the free, Abelian case, see e.g.~\cite{Witten:2007ct}. There are a number of arguments supporting this belief, some of which are explained in other contributions to this volume, see in particular~\cite{contrib:lambert}. 

The most commonly voiced (but also weakest) one is the argument that the (2,0)-theory comes with neither dimensionful nor dimensionless parameters and therefore cannot possess a classical limit or a Lagrangian. A precise formulation of this point was given in~\cite{Bekaert:9909094,Bekaert:2000qx}, where it was shown that the free Abelian (2,0)-theory does not allow for continuous deformations to an interacting field theory. From a wider perspective, however, this is in fact expected. Recall that M2-brane models are Chern--Simons matter theories which by their very nature come with a discrete coupling constant. We expect, and indeed shall see, that M5-brane models are higher analogues of Chern--Simons matter theories and thus also require a discretization of the coupling constant.

A second common argument is that a parallel transport of higher-dimensional objects such as the self-dual strings can only be defined in the Abelian case. This statement is based on the famous Eckmann--Hilton argument, cf.~\cite{Saemann:2016sis} for more details. It is certainly true, if we work with ordinary gauge groups. Higher-dimensional parallel transport, however, is intrinsically linked to higher-dimensional algebra. Correspondingly, we should work with higher or categorified gauge groups, which allow for a consistent general parallel transport of higher-dimen\-sional objects. The resulting gauge theories are known as ``higher gauge theories.''

\section{Higher gauge theories}

\subsection{Motivation}

A first, crude motivation for considering higher gauge theories is the following one. Many interesting features of string theory can be studied within ordinary gauge theories. These arise in string theory when we turn off gravity and consider the dynamics of the endpoints of strings on D-branes. The connections or gauge potentials of the gauge theories then capture the parallel transport of these endpoints. Lifting this picture to M-theory, the strings and D-branes get replaced by M2- and M5-branes, and we are now interested in a parallel transport of the one-dimensional boundaries of the M2-branes on the M5-branes. To render the resulting connective structures interacting, we have to switch to higher-dimensional algebra and ultimately higher gauge theory, as mentioned above.

More concretely, we know that the field content of the $(2,0)$-theory comprises the $\CN=(2,0)$ tensor multiplet, which contains a 2-form potential $B$. In the type IIB interpretation of the $(2,0)$-theory, this is the usual Kalb--Ramond field of string theory, which we know is part of the connective structure of an Abelian gerbe~\cite{Gawedzki:1987ak,Freed:1999vc}. The latter is the mathematical structure underlying Abelian higher gauge theory. For the interacting non-Abelian generalization, we are therefore led to considering connective structures on non-Abelian gerbes or higher principal bundles, for which a complete mathematical theory exists and which underlie higher gauge theories.

We also note that the observables of the $(2,0)$-theory are Wilson surfaces as opposed to the Wilson lines of ordinary gauge theories, as argued in~\cite{Ganor:1996nf}.

Altogether, we have rather clear indication that the $(2,0)$-theory is a higher gauge theory. Whether this higher gauge theory admits a classical description in terms of equations of motion or even an action, is a different question. Also, we would expect that significant insights into the (2,0)-theory can be won by quantizing a classical higher gauge theory, even if this theory is not precisely the classical description of the $(2,0)$-theory.

It is therefore our goal to try to construct a six-dimen\-sional classically superconformal field theory in the form of a higher gauge theory which shares as many features as possible with the $(2,0)$-theory.

The transition from ordinary to higher gauge theory is made by replacing the kinematical data of gauge theory, in particular gauge groups, connections on principal fibre bundles, and sections of associated vector bundles by categorified analogues. Here, we restrict ourselves to the case of local higher gauge theory with infinitesimal symmetries, which means that we merely have to categorify gauge Lie algebras and local gauge potentials.

\subsection{$L_\infty$-algebras}

A very nice and useful description of categorified Lie algebras is given in terms of {\em strong homotopy Lie algebras}, or {\em $L_\infty$-algebras} for short. These were first introduced in the context of closed string field theory~\cite{Zwiebach:1992ie} as analogues of Stasheff's homotopy associative algebras~\cite{Stasheff:1963aa,Stasheff:1963ab}, or $A_\infty$-algebras for short. The fact that Lie 2-algebras are indeed categorically equivalent to a particular subset of $L_\infty$-algebras can be found in~\cite{Baez:2003aa}. Strong homotopy Lie algebras are also one of the first higher structures one should learn, both because of their importance and their simplicity. For a very detailed review, see e.g.~\cite{Jurco:2018sby}.

Recall that a Lie algebra is a vector space $\frg$ with a bilinear, antisymmetric product $[-,-]:\frg^{\wedge 2}\rightarrow \frg$ which satisfies the Jacobi identity
\begin{equation}\label{eq:jacobi}
 [[a_1,a_2],a_3]-[[a_1,a_3],a_2]+[[a_2,a_3],a_1]=0~.
\end{equation}

Mathematical structures in the Bourbaki style consist of sets, structure maps and structure equations satisfied by the structure maps. Categorifying a mathematical structure now means in particular to replace the underlying sets by sets of objects, sets of morphisms, sets of morphisms between morphisms and so on. The level of morphisms gives rise to a grading, and an $L_\infty$-algebra therefore has an underlying graded vector space $\sL=\oplus_{k} \sL_k$ with $k\leq 0$ and $-k$ the level of the morphism. The definition of $L_\infty$-algebra, however, extends to $k\in \RZ$. 

When an $L_\infty$-algebra is non-trivial only in particular degrees, we say that it is {\em concentrated} in these degrees. An $L_\infty$-algebra concentrated in degrees $-n+1,\dots,0$ corresponds to an $n$-fold categorification of a Lie algebra (with morphisms up to degree~$n-1$) and these are called {\em Lie $n$-algebras}. We shall also use this term for the corresponding $L_\infty$-algebras.

The structure map of a Lie algebra, namely the Lie bracket $[-,-]$, is then replaced by a higher structure functor, which satisfies the structure identity~\eqref{eq:jacobi} only up to an isomorphism. This isomorphism is encoded in additional maps, which have to satisfy a number of coherence axioms. In the $L_\infty$-algebra picture, the structure functors and the additional maps turn into totally antisymmetric multilinear products $\mu_i:\sL^{\wedge i}\rightarrow \sL$ of degree~$2-i$ and the weakened Jacobi identity as well as all higher coherence axioms can be summed up by the higher or homotopy Jacobi relation
\begin{equation}\label{eq:hom_rel}
\begin{aligned}
\sum\limits_{i+j=n} &\sum\limits_\sigma (-1)^{j} \chi(\sigma; \ell_1,\dots,\ell_n)\times\\
&\times \mu_{j+1}(\mu_i(\ell_{\sigma(1)},\dots,\ell_{\sigma(i)}),\ell_{\sigma(i+1)},\dots,\ell_{\sigma(n)}) = 0
\end{aligned}
\end{equation}
for all $n\in\NN$ and $\ell_1,\dots,\ell_n\in \sL$. Here, the second sum runs over all $(i,j)$-``unshuffles,'' i.e.~permutations of $1,\dots,n$ for which the first $i$ and last $j$ numbers are ordered. Moreover, $\chi(\sigma; \ell_1,\dots,\ell_n)$ denotes the Koszul sign obtained by graded antisymmetric permutation of objects. 

Let us give a few examples. First, we have the trivial $L_\infty$-algebra with $\sL_k=*$ for all $k$, where $*=\{0\}$ is the $0$-dimensional vector space. All products $\mu_i$ are trivial and map their arguments to the null vector in $\sL=\oplus_k \sL_k$.

Second, any Lie algebra $\frg$ gives rise to an $L_\infty$-algebra if we set $\sL_0=\frg$ and $\sL_k=*$ for $k\neq 0$. The only non-trivial product here is $\mu_2$ on $\sL_0^{\wedge 2}$, which equals the Lie bracket $[-,-]$.

Third, any Lie algebra $\frg$ together with a representation $\rho:\frg\mapsto \sEnd(V)$, where $V$ is the representation space, forms an $L_\infty$-algebra concentrated in degrees~$0$ and $-1$ with $\sL_0=\frg$, $\sL_{-1}=V$. The non-trivial products are
\begin{equation}
\begin{aligned}
 \mu_2(a_1,a_2)&=[a_1,a_2]~,\\
 \mu_2(a_1,v)&=\rho(a_1)v
\end{aligned}
\end{equation}
for all $a_1,a_2\in \frg$ and $v\in V$.

The last example is interesting in our context as the 3-Lie algebras\footnote{not to be confused with Lie 3-algebras, which are essentially $L_\infty$-algebras concentrated in degrees $-2$, $-1$ and $0$} relevant in M2-brane models are, in fact, of this type~\cite{Palmer:2012ya}, see also~\cite{Saemann:2016sis}. Furthermore, one can in principle regard M2-brane models as higher gauge theories~\cite{Palmer:2013ena}. This is again encouraging, since we expect some relation between the (2,0)-theory and the M2-brane models.

Fourth, the de Rham complex on some manifold $M$ forms an $L_\infty$-algebra with $\sL_k=\Omega^k(M)$ and the only non-trivial bracket being the de Rham differential $\mu_1=\dd$. Note that for a $d$-dimensional manifold $M$, this $L_\infty$-algebra $\sL$ is concentrated in degrees~$0,\dots,d$ contrary to the Lie $n$-algebras, which are concentrated in non-positive degrees.

The above examples are all {\em strict $L_\infty$-algebras}, that is $L_\infty$-algebras with $\mu_i$ trivial for $i\geq 3$. Such $L_\infty$-algebras are the same as differential graded Lie algebras. To go beyond this class of examples, we need to specify higher products and this is most readily done in terms of cocycles. An example crucial for our discussion will be the following one~\cite{Baez:2003aa}. Let $\frg$ be a semisimple Lie algebra. Its Cartan--Killing form $(-,-)$ is non-degenerate and gives rise to the Lie algebra 3-cocycle $(-,[-,-])$. We then have the $L_\infty$-algebra
\begin{subequations}\label{eq:string_Lie_2}
\begin{equation}
 \sL=\sL_{-1}\oplus \sL_0~,~~~\sL_0=\frg~,~~~\sL_{-1}=\FR\\
\end{equation}
with non-trivial products
\begin{equation}
\begin{aligned}
 \mu_2(a_1,a_2)&=[a_1,a_2]~,\\
 \mu_3(a_1,a_2,a_3)&=(a_1,[a_2,a_3])
\end{aligned}
\end{equation}
\end{subequations}
for $a_{1,2,3}\in \sL_0$. In the special case $\frg=\aspin(n)$, this is often called the {\em string Lie 2-algebra}; we will use this term in general and write $\astring_{\rm sk}(\frg)$ for the above $L_\infty$-algebra. The subscript sk indicates that this $L_\infty$-algebra is {\em skeletal}, i.e.~$\mu_1=0$. This example will turn out be the foundation of our field theory later on.

Given a closed, non-degenerate 3-form on a manifold, i.e.~a {\em multisymplectic form}, one can define an $L_\infty$-algebra on the vector space of differential 1-forms and functions, which is some higher version of the Poisson algebra on a symplectic manifold~\cite{Baez:2008bu}. In the cases where the multisymplectic manifold is a compact simple Lie group $\sG$ endowed with the canonical 3-form, we can restrict the resulting higher Poisson algebra to the string Lie 2-algebra $\astring(\sLie(\sG))$~\cite{Baez:2009:aa}.

The last two examples are special cases of a general construction mechanism of $L_\infty$-algebras in which $L_\infty$-algebras concentrated in fewer degrees are extended to ones concentrated in more degrees via cocycle data, cf.~also~\cite{Fiorenza:2013nha}.

We conclude that a higher gauge theory, and in particular the field theory we wish to construct should have a gauge $L_\infty$-algebra replacing an ordinary gauge Lie algebra. Recall from above that $L_\infty$-algebras were first defined in the context of closed string field theory~\cite{Zwiebach:1992ie} and indeed govern the field content and the symmetries of this theory. It is certainly encouraging to find that the mathematical structure we are led to use in a description of the $(2,0)$-theory is already fundamentally contained in string field theory.

\subsection{Quasi-isomorphisms}

Due to their higher categorical nature, maps between $L_\infty$-algebras are now much richer than maps between Lie algebras. Naively, one might think that morphisms between $L_\infty$-algebras $(\sL,\mu_i)$ and $(\sL',\mu'_i)$ should be given by grade preserving linear maps $\phi:\sL\rightarrow \sL'$ compatible with the higher products in the sense that 
\begin{equation}
 \phi(\mu_i(\ell_1,\dots,\ell_i))=\mu'_i(\phi(\ell_1),\dots,\phi(\ell_i))
\end{equation}
for all $i\in \NN$. These, however, are just the {\em strict} morphisms. More general morphisms are readily derived from the dual description of $L_\infty$-algebras in terms of differential graded commutative algebras, see~\cite{Lada:1994mn} or also~\cite{Jurco:2018sby} for more details. The end result is that a morphism of $L_\infty$-algebras $\phi:\sL\rightarrow \sL'$ corresponds to a set of maps $\phi_i:\sL^{\wedge i}\rightarrow \sL'$ of degree~$1-i$ such that
\begin{subequations}\label{eq:LMor}
\begin{equation}
\begin{aligned}
   &\sum_{j+k=i}\sum_{\sigma\in {\rm Sh}(j;i)}~(-1)^{ij}\chi(\sigma;\ell_1,\ldots,\ell_i)\times\\
   &\kern.5cm\times \phi_{k+1}(\mu_j(\ell_{\sigma(1)},\dots,\ell_{\sigma(j)}),\ell_{\sigma(j+1)},\dots ,\ell_{\sigma(i)})=\\
   &=\sum_{j=1}^i\frac{1}{j!} \sum_{k_1+\cdots+k_j=i}\sum_{\sigma\in{\rm Sh}(k_1,\ldots,k_{j-1};i)}\times\\
   &\kern.5cm\times\chi(\sigma;\ell_1,\ldots,\ell_i)\zeta(\sigma;\ell_1,\ldots,\ell_i)\,\times\\
   &\kern1cm\times \mu'_j\Big(\phi_{k_1}\big(\ell_{\sigma(1)},\ldots,\ell_{\sigma(k_1)}\big),\ldots,\\
   &\kern2.5cm \phi_{k_j}\big(\ell_{\sigma(k_1+\cdots+k_{j-1}+1)},\ldots,\ell_{\sigma(i)}\big)\Big)~,
\end{aligned}
\end{equation}
with the sign
\begin{equation}
\begin{aligned}
 &\zeta(\sigma;\ell_1,\ldots,\ell_i):=\\
 &\kern.5cm:=(-1)^{\sum_{1\leq m<n\leq j}k_mk_n+\sum_{m=1}^{j-1}k_m(j-m)}\times\\
 &\kern1cm\times(-1)^{\sum_{m=2}^j(1-k_m)\sum_{k=1}^{k_1+\cdots+k_{m-1}}|\ell_{\sigma(k)}|}~.
 \end{aligned}
\end{equation}
\end{subequations}
Strict morphisms are recovered if the $\phi_i$ for $i>1$ are trivial.

We note that $\phi_1$ is still a chain map of the complexes $(\sL,\mu_1)$ and $(\sL',\mu'_1)$. On chain complexes, we have the notion of {\em quasi-isomorphisms}. A quasi-isomorphism between two chain complexes $(\sL,\mu_1)$ and $(\sL',\mu'_1)$ can be encoded in the following maps:
\begin{equation}
 \myxymatrix{
\sL \ar@/^2ex/[r]^{\phi} &
\sL'  
  \ar@/^2ex/[l]^{\psi}
}~,~~~\begin{array}{c}
\tau_{\sL}:\psi\circ \phi \cong \id_{\sL}\\
\tau_{\sL'}:\phi\circ \psi \cong \id_{\sL'}
\end{array}~~,
\end{equation}
where $\psi$ and $\phi$ are chain maps and $\tau_{\sL}$ and $\tau_{\sL'}$ are chain homotopies. Equivalently, a quasi-isomorphism is given by a chain map $\phi:(\sL,\mu_1)\rightarrow (\sL',\mu'_1)$, which induces an isomorphism\footnote{Note that chain maps descend to cohomology.} between the cohomologies $H^\bullet_{\mu_1}(\sL)$ and $H^\bullet_{\mu'_1}(\sL')$.

This picture can be generalized to $L_\infty$-algebras. We call two $L_\infty$-algebras $(\sL,\mu_i)$ and $(\sL,\mu'_i)$ {\em quasi-isomorphic}, if there is a morphism $\phi:\sL\rightarrow \sL'$ such that $\phi_1$ induces an isomorphism between the cohomologies $H^\bullet_{\mu_1}(\sL)$ and $H^\bullet_{\mu'_1}(\sL')$. 

In the case of an $L_\infty$-algebra $\sL=\sL_0$ concentrated in degree~$0$, i.e.~an ordinary Lie algebra, quasi-isomorphisms are ordinary isomorphisms since $\mu_1$ is trivial and the cohomology of $\sL$ is just $\sL$. As a non-trivial example, consider an $L_\infty$-algebra $\sL=\sL_{-1}\oplus \sL_0$ with $\sL_{-1}=\sL_0=V$ and $\mu_1|_{\sL_{-1}}=\id$. Since there is an obvious chain map from $\sL$ to the trivial $L_\infty$-algebra and because the cohomology of $\sL$ is trivial, $\sL$ is quasi-isomorphic to the trivial $L_\infty$-algebra.

A crucial result is now the {\em minimal model theorem}~\cite{kadeishvili1982algebraic}, which states that any $L_\infty$-algebra is quasi-isomorphic to a skeletal $L_\infty$-algebra. Since the complex $(\sL,\mu_1)$ of such an $L_\infty$-algebra $\sL$ equals its cohomology $H^\bullet_{\mu_1}(\sL)$, such an $L_\infty$-algebra is indeed a minimal representative of its quasi-isomorphism class and therefore called a {\em minimal model}. One can think of such an $L_\infty$-algebra as one where all redundancy has been factored out.

There are a number of reasons for why quasi-isomor\-phisms are indeed the correct notion of isomorphisms, chief of which for us is perhaps the point that higher principal bundles are only properly defined modulo equivalences of their higher structure groups. The latter equivalence corresponds to quasi-isomorphisms of $L_\infty$-algebras at the level of higher structure Lie algebras.

Altogether, we conclude here that a consistent higher gauge theory such as the one we want to construct should respect quasi-isomorphisms. Physics should not care about which concrete model of the gauge $L_\infty$-algebra we choose.\footnote{This holds true modulo a well-understood subtlety in category theory. The model should be ``large enough'' to allow for enough morphisms ``out of it.'' A simple and related example is the choice of cover of a manifold. While all covers of some manifold $M$ are equivalent in a precise sense (their \v Cech groupoids are Morita equivalent), we need a cover that is fine enough to describe all principal bundles over $M$.} This will prove to be a strong consistency check of our theory which is completely invisible from a perspective ignoring higher structures.

\subsection{Kinematical data of higher gauge theories}

Given an $L_\infty$-algebra $\sL$, we need to construct the kinematical data of a corresponding higher gauge theory next. That is, we have to specify notions of gauge potentials, curvatures, gauge transformations, Bianchi identities and, ideally, topological invariants. There are a number of ways of doing this, generalizing the usual definition of connections on principal bundles.

Here we shall take a slight shortcut as done e.g.~in~\cite{Jurco:2014mva}. We note that $L_\infty$-algebras are generalizations of differential graded Lie algebras. For the latter, we have the Maurer--Cartan equation, which naturally generalizes to $L_\infty$-algebras, see e.g.~\cite{Jurco:2018sby} for a detailed review. A gauge potential is here an element $a\in \sL_1$, and we call it a {\em homotopy Maurer--Cartan element} if it satisfies 
the {\em homotopy Maurer--Cartan equation}
\begin{equation}
\begin{aligned}
 f&:=\mu_1(a)+\tfrac12 \mu_2(a,a)+\tfrac{1}{3!}\mu_3(a,a,a)+\cdots\\
 &\phantom{:}=\sum_{i=1}^\infty \tfrac{1}{i!} \mu_i(a,\dots,a)=0~.
\end{aligned}
\end{equation}
The element $f\in \sL_2$ is the curvature of $a$ and it satisfies the Bianchi identity 
\begin{equation}
 \sum_{i=0}^\infty \frac{(-1)^i}{i!} \mu_{i+1}(f,a,\dots,a)=0~.
\end{equation}
The gauge transformations are given by 
\begin{equation}
 \delta a=\sum_{i=0}^\infty \tfrac{1}{i!}\mu_{i+1}(a,\dots,a,\alpha)
\end{equation}
for $\alpha$ an element of $\sL_0$. Thus, any $L_\infty$-algebra comes naturally with the kinematical data of a (very abstract) gauge theory.

For the $L_\infty$-algebra given by the de Rham complex $\sL=\Omega^\bullet(M)$ of some manifold $M$, we obtain the kinematical data for the topologically trivial sector of an ordinary Abelian gauge theory. Note that we can shift the degree of the de Rham complex: $\sL_\bullet=\Omega^{\bullet+k}(M)$, and for $k>1$, we obtain the kinematical data for the topologically trivial sector of higher Abelian gauge theories.

The restriction to the topologically trivial sector is due to our restriction to global differential forms and therefore topologically trivial higher principal bundles. The restriction to Abelian gauge theory arises from restricting to ordinary, real-valued differential forms. We can, however, consider non-Abelian examples by tensoring differential forms by a gauge $L_\infty$-algebra. This is possible since the tensor product of a differential graded commutative algebra and an $L_\infty$-algebra $\sL$ carries a natural $L_\infty$-algebra structure. Here, we are interested in the tensor product
\begin{subequations}
 \begin{equation}
 \hat \sL=\Omega^\bullet(M)\otimes \sL=\bigoplus_{k\in\RZ} \hat \sL_{k}=\bigoplus_{k\in\RZ} \bigoplus_{i-j=k} \Omega^i(M)\otimes \sL_j~,
\end{equation}
which carries the $L_\infty$-algebra structure
\begin{equation}\label{eq:tensor_mui}
\begin{aligned}
 \hat \mu_1(\omega_1\otimes \ell_1)&\coloneqq \dd \omega_1 \otimes \ell_1\pm\omega_1\otimes \mu_1(\ell_1)~,\\
 \hat \mu_i(\omega_1\otimes \ell_1,\dots,\omega_i\otimes \ell_i)&\coloneqq 
 \pm (\omega_1\wedge \dots \wedge \omega_i)\otimes \mu_i(\ell_1,\dots,\ell_i)~,
\end{aligned}
\end{equation}
\end{subequations}
where $\omega_j$ and $\ell_j$ are homogeneous elements in $\Omega^\bullet(M)$ and $\sL$ (with obvious linear continuation to inhomogeneous elements) and the signs arise from permuting differential forms of odd degree with odd elements $\ell_j$ or a product $\mu_i$ with $i$ odd.

Let us now consider some manifold $M$ together with a gauge Lie 2-algebra, i.e.~an $L_\infty$-algebra concentrated in degrees~$0$ and $-1$: $\sL=\sL_{-1}\oplus \sL_0$. The kinematical data of $\hat \sL=\Omega^\bullet(M)\otimes \sL$ is then that of a higher connection on a topologically trivial principal 2-bundle with structure Lie 2-algebra $\sL$. Concretely, we have a gauge potential
\begin{subequations}\label{eq:higher_gauge_2-Lie}
\begin{equation}
 a=A+B~,~~~A\in \Omega^1(M)\otimes \sL_0\eand B\in \Omega^2(M)\otimes \sL_{-1}
\end{equation}
with curvature 
\begin{equation}\label{eq:curvature_components_2-Lie}
\begin{aligned}
 f&=\CF+H~,\\
 &\CF=\dd A+\tfrac12 \mu_2(A,A)+\mu_1(B)~~~\in \Omega^2(M)\otimes \sL_0~,\\
 &H=\dd B+\mu_2(A,B)+\tfrac{1}{3!}\mu_3(A,A,A)~~~\in \Omega^3(M)\otimes \sL_{-1}~.
\end{aligned}
\end{equation}
The 2-form $\CF$ is also called {\em fake curvature} in the context of non-Abelian gerbes. The components of the curvature satisfy each a Bianchi identity and gauge transformations of the components of the gauge potential and the curvature are readily computed. We merely note that the latter transform according to
\begin{equation}\label{eq:gauge_trafos_curvatures_2-Lie}
\begin{aligned}
 \delta \CF&=\mu_2(\CF,\alpha)~,\\
 \delta H&=\mu_2(H,\alpha)+\mu_2(\CF,\Lambda)-\mu_3(\CF,A,\alpha)~,
\end{aligned}
\end{equation}
where $\alpha\in\Omega^0(M)\otimes\sL_0$ and $\Lambda\in \Omega^1(M)\otimes \sL_{-1}$ are the gauge parameters.
\end{subequations}

In this manner, we can in principle compute the kinematical data for the topologically trivial sector of any higher gauge theory.

\subsection{Fake curvature and redefinition of curvatures}

The above definitions are relatively straightforward, given some familiarity with $L_\infty$-algebras. This begs the questions why these structures have not been applied successfully before in the context of the (2,0)-theory. We shall explain a potential reason in the following.

If we compute the commutator of two gauge transformations of a gauge potential $a$ in an $L_\infty$-algebra, we obtain 
\begin{subequations}\label{eq:CommutatorGT}
\begin{equation}
  [\delta_{ \alpha_0},\delta_{ \alpha'_0}]a\ =\ \delta_{ \alpha''_0}a+\sum_{i\geq0}\frac{1}{i!}(-1)^{i}\mu_{i+3}(f,a,\ldots,a, \alpha_0, \alpha'_0)
\end{equation}
for $\alpha_0,\alpha'_0\in \sL_0$ with
\begin{equation}
   \alpha''_0\ :=\ \sum_{i\geq0}\frac{1}{i!}\mu_{i+2}(a,\ldots,a, \alpha_0, \alpha'_0)~.
\end{equation}
\end{subequations}
We see that gauge transformations only close if $f=0$ wich in the concrete example of the higher gauge theory~\eqref{eq:higher_gauge_2-Lie} amounts to $\CF=0$. The necessity for imposing $\CF=0$ is also seen from other perspectives. If we wanted to have self-duality of the 3-form $H$ from~\eqref{eq:curvature_components_2-Lie} as a consistent equation of motion, invariance of $H=*H$ under the gauge transformations~\eqref{eq:gauge_trafos_curvatures_2-Lie} clearly requires $\CF=0$ (or restricting gauge transformations to those with $\Lambda=0$, which is very problematic). Moreover, it has been shown that for the kinematical data~\eqref{eq:higher_gauge_2-Lie}, a parallel transport along surfaces is only reparametrization invariant, if $\CF=0$~\cite{Baez:2004in}. Vanishing of the fake curvature also arises naturally in twistor descriptions based on conventional higher holomorphic principal bundles~\cite{Saemann:2012uq,Saemann:2013pca,Jurco:2014mva,Jurco:2016qwv}.

One could now consider the condition $\CF=0$ as an equation of motion and postulate that the above relation~\eqref{eq:CommutatorGT} indicates an ``open symmetry,'' which closes only up to equations of motion. This interpretation, however, leads to another problem. If we assume that physics should not care about the model of the gauge $L_\infty$-algebra, we can replace our gauge $L_\infty$-algebra by its minimal model $\sL'=\sL'_0\oplus \sL'_{-1}$, for which $\mu'_1=0$ and $\mu'_2$, restricted to $\sL_0'$, is a Lie bracket. The equation $\CF=0$ then amounts to
\begin{equation}
 F=\dd A+\tfrac12 [A,A]=0~,
\end{equation}
and we can gauge away $A$, at least locally. The curvature $H$ then reduces to the Abelian one. It seems to be essentially this argument which led to claims in the literature that non-Abelian gerbes are still essentially Abelian (even though we are not aware of it being stated anywhere as clearly as above).

We thus conclude that for our higher gauge theory approximating the $(2,0)$-theory, the above derived kinematical data as it stands is {\em not} suitable. A small modification, however, allows us to remedy the situation.

We note that our construction of higher gauge structures from homotopy Maurer--Cartan theory is a vast generalization of Chern--Simons theory, where $f=0$ or, for a higher gauge theory based on a Lie 2-algebra, $\CF=0$ would indeed be part of the equations of motion. There are therefore no issues in defining higher Chern--Simons theories along the way we indicated.

It seems that while the kinematical data of Chern--Simons theory and Yang--Mills theory agree for ordinary gauge theory, the generalization to higher gauge theory now depends on whether we are interested in a theory with vanishing curvature, as higher Chern--Simons theory, or in a theory with not necessarily vanishing curvature, as e.g.~the (2,0)-theory. This seems to be an important observation which is not contained in the literature, at least not to our knowledge.

In the case of theories with not necessarily vanishing curvatures, one sees this problem rather generically. The construction of topological invariants as suggested in~\cite{Sati:2008eg} is {\em not} compatible with quasi-isomorphisms. More precisely, the topological invariants arise from the invariant polynomials of an $L_\infty$-algebra. The latter form a differential graded algebra which is dual to an $L_\infty$-algebra. Given two quasi-isomorphic $L_\infty$-algebras, their $L_\infty$-algebras of invariant polynomials are not necessarily quasi-isomorphic, which violates our principle of consistency with quasi-isomorphism. Again, in the case of higher Chern--Simons theories, the topological invariants become meaningless, which renders the problem of incompatibility with quasi-isomorphisms irrelevant.

The way out is familiar to physicists from heterotic supergravity and to mathematicians from the context of {\em string structures}: we need to modify our notion of curvature. For reasons which go beyond the scope of this review, this modification also alters the gauge transformations. A detailed motivation and discussion of this modification is found e.g.~in~\cite{contrib:schmidt}.

\subsection{String structures}

In heterotic supergravity, it is known that in order to couple a potential 1-form $A$ taking values in a non-Abelian Lie algebra $\frg$ to the Kalb--Ramond $B$-field 2-form, one has to modify the expression for the 3-form curvature $\dd B$. Explicitly, one defines 
\begin{subequations}\label{eq:string_structure_skeletal}
\begin{equation}
 H:=\dd B+{\rm cs}(A)-{\rm cs}(\omega)~,
\end{equation}
where ${\rm cs}(-)$ is the (trace over the) Chern--Simons form and $\omega$ is the spin connection $1$-form, taking values in $\aspin(1,9)$~\cite{Bergshoeff:1981um,Chapline:1982ww}. The above 3-form curvature is complemented by the 2-form curvatures
\begin{equation}
 \begin{aligned}
    \CF_\omega&=\dd \omega+\tfrac12[\omega,\omega]~,\\
    \CF_A&=\dd A+\tfrac12[A,A]~.
 \end{aligned}
\end{equation}
The infinitesimal gauge transformation of the involved potential forms $\omega$, $A$, and $B$ reads as
\begin{equation}
\begin{aligned}
 \delta \omega &= \dd \lambda + [\omega,\lambda]~,\\
 \delta A &= \dd\alpha + [A,\alpha]~,\\
 \delta B &= \dd \Lambda +(\lambda,\dd \omega)-(\alpha,\dd A)
\end{aligned}
\end{equation}
with $\lambda \in \Omega^0(M)\otimes \aspin(1,9)$, $\alpha\in \Omega^0(M)\otimes \frg$ and $\Lambda \in \Omega^1(M)$ the gauge parameters. The resulting Bianchi identities are
\pagebreak
\begin{equation}\label{eq:gg_Bianchi}
\begin{aligned}
 \dd \CF_\omega+[\omega,\CF_\omega]&=0~,\\
 \dd \CF_A+ [A,\CF_A]&=0~,\\
 \dd H-(F_\omega,F_\omega)+(F_A,F_A)&=0~,
\end{aligned}
\end{equation}
\end{subequations}
and we recognize the {\em Green--Schwarz anomaly cancellation condition}, which guarantees that the gauge and the gravitational anomalies cancel each other~\cite{Green:1984sg} in the last equation.

This ad-hoc construction can also be motivated and obtained from a purely mathematical perspective. Recall that the closed strings of string theory lead us to considering the loop space $\CL M$ of our space-time $M$. Roughly speaking, ordinary structures on loop spaces correspond to once categorified ones or rather once degree shifted ones on the original space-time. For example, loop space $\CL M$ should be regarded orientable, if $M$ carries a spin structure~\cite{Witten:1987cg,AST_1985__131__43_0}. As argued then in~\cite{Killingback:1986rd}, a spin structure on $\CL M$ corresponds to a {\em string structure} on $M$. 

Formally, a string structure on a principal $\sSpin(n)$-bundle $P$ is a lift of the structure group to a central extension. This central extension is only possible if the first fractional Pontryagin class of $P$ vanishes: $\tfrac12 p_1(P)=0$~\cite{mclaughlin1992orientation,Stolz:2004aa}. This class is the characteristic class of the Chern--Simons 2-gerbe of $P$~\cite{Carey:0410013}, and a string structure can be regarded as a principal $\sSpin(n)$-bundle $P$ with a trivialization of the Chern--Simons gerbe~\cite{Waldorf:2009uf}. The latter is essentially the content of the last equation of~\eqref{eq:gg_Bianchi}: the 3-form $H$ is the trivialization of $\tfrac12 p_1(P)$, with the inclusion of the Pontryagin class of the tangent bundle.

This can be formulated even more directly in the context of higher gauge theory, where it boils down to rendering the definition of invariant polynomials compatible with quasi-isomorphisms, cf.~the discussion in the last subsection. Again, the details are beyond the scope of our discussion, but can be found in~\cite{Sati:2009ic,Fiorenza:2010mh}, see also~\cite{Saemann:2017rjm} and~\cite{contrib:schmidt}. The result is the modification of the usual definition of the kinematical data of higher gauge theory for the string Lie 2-algebra~\eqref{eq:string_Lie_2} to the modified curvatures and gauge transformations of~\eqref{eq:string_structure_skeletal}. 

For string structures, the gauge transformations close directly; in particular $H$ itself is gauge invariant, and there is no issue with covariance of the field equation $H=*H$ for $M=\FR^{1,5}$.

We conclude that there is at least one example of kinematical data that allows for non-trivial non-Abelian extensions, namely the data of differential string structures. Moreover, their intrinsic connection to string theory via spin bundles over loop space is again encouraging with regard to a theory of self-dual strings.

\section{Self-dual strings}

We saw above that the explicit construction of the kinematical data highly depends on the chosen higher structure group, a phenomenon which slightly resembles the situation for M2-brane models, where the gauge group determined the maximally available amount of supersymmetry. In the next step, we should therefore identify potentially interesting higher gauge groups. To this end, we can consider concrete examples in ordinary gauge theory and try to translate them to higher analogues.

The simplest interesting example of a non-Abelian gauge group is certainly $\sSU(2)$. The underlying manifold is $S^3$, which is itself a principal circle bundle over $S^2$, known as the {\em Hopf fibration}. The connection on this bundle describes the gauge potential of a magnetic monopole of charge~$1$ located at the center of the sphere. We may find higher analogues of~$\sSU(2)$ by considering higher analogues of monopoles.

\subsection{Categorified monopoles}

The Dirac monopole in $\FR^3$ arises from completing Max\-well's equation to an electric-magnetic dually symmetric form by inserting magnetic sources. Since this breaks the Bianchi identity $\dd F=0$, the Poincar\'e lemma does not apply at the locations of monopoles and thus the electromagnetic field does not admit a gauge potential there. If the magnetic source is located at a single point in $\FR^3$, we can describe the configuration in terms of a connection on a principal $\sU(1)$-bundle over $S^2$ as mentioned above and the charge of the monopole is the first Chern number of this bundle. More concretely, we have an additional Higgs field $\phi$, which is a section of the associated line bundle with respect to the fundamental representation satisfying the Bogomolny equation $F=*\dd \phi$.

The Dirac monopole has a string theory interpretation in terms of a D1-brane ending on a D3-brane with $\FR^3$ being the spatial part of the world-volume of the D3-brane and the location of the monopole being the endpoint of the D1-brane. This configuration is readily lifted to M-theory where an M2-brane ends on an M5-brane in a self-dual string parallel to both of the M-branes worldvolumes. This M-brane configuration is described by a dimensional reduction of the self-dual 3-form $H=\dd B$ to the four spatial dimensions of the M5-brane's worldvolume perpendicular to the self-dual string. That is, we have an Abelian 2-form $B$ on $\FR^4$ together with an Abelian Higgs field satisfying the equation $H=\dd B=*\dd \phi$~\cite{Howe:1997ue}. Geometrically, the $B$-field is part of the connective structure of a gerbe. If we again assume a spherical symmetric situation due to self-dual strings whose boundaries are at the same point in $\FR^4$, we can describe this configuration by a gerbe over $S^3$ whose center is the location of the self-dual strings. In this sense, self-dual strings are categorified monopoles.

We are now interested in a non-Abelian generalization of the above two pictures. In the monopole case, there is the 't Hooft--Polyakov monopole with underlying gauge group $\sSU(2)$~\cite{Hooft:1974:276,Polyakov:1974ek}. This monopole extends to all of $\FR^3$, and the corresponding principal $\sSU(2)$-bundle over~$\FR^3$ is necessarily trivial. Asymptotically, however, the bundle still resembles the Dirac monopole. More precisely, there is an asymptotic morphism of principal bundles from that of the Dirac monopole of charge~1 to the trivial principal $\sSU(2)$-bundle of the 't Hooft--Polyakov monopole at the 2-sphere at infinity. The key to this morphism is the fact that the total space of the former becomes the structure group of the latter.

This motivates us to look for a 2-group structure on the total categorified space of the fundamental gerbe\footnote{Gerbes can be regarded as central groupoid extensions, so that the total categorified space of a gerbe is simply a Lie groupoid.} of charge~1 over $S^3\cong \sSpin(3)$. Such a 2-group structure indeed exists, and it forms in fact a 2-group model of the string group $\sString(3)$, which differentiates to the string Lie 2-algebra~\eqref{eq:string_Lie_2} for $\frg=\aspin(3)$ as shown in~\cite{Demessie:2016ieh}. As we know from above, higher gauge theory with this underlying Lie 2-algebra allows for a modification that admits closed gauge transformations and gauge covariant field equations independent of whether the fake curvature vanishes.

There are, in fact, many more reasons for choosing the string Lie 2-algebra as a higher gauge algebra, and we refer to the papers~\cite{Saemann:2017rjm,Saemann:2017zpd} for further details.

\subsection{More on the string Lie 2-algebra}

Let us have a brief, more detailed look at the string Lie 2-algebra. The form of the string Lie 2-group which was differentiated in~\cite{Demessie:2016ieh} is, in fact, quite involved. However, there is a strict version\footnote{Recall that 'strict' means $\mu_i=0$ for $i>3$.} of the string Lie 2-algebra, which is quasi-isomorphic to the skeletal model~$\astring_{\rm sk}(\frg)$, and which is readily integrated:
\begin{subequations}\label{eq:loop_model}
\begin{equation}
 \begin{aligned}
  \astring_{\Omega}(\frg)&=\astring_{\Omega,-1}(\frg)\oplus \astring_{\Omega,0}(\frg)~,\\
  &\astring_{\Omega,-1}(\frg)= \hat \Omega \frg\cong \Omega \frg\oplus \FR~,\\
  &\astring_{\Omega,0}(\frg)= P_0 \frg~,
 \end{aligned}
\end{equation}
where $P_0 \frg$ and $\Omega \frg$ are based path and loop spaces of $\frg$ and $\hat \Omega\frg$ is the Lie algebra version of the level-1 Kac--Moody central extension of $\Omega \frg$. The differential 
\begin{equation}
 \mu_1:\hat \Omega\frg \rightarrow P_0 \frg
\end{equation}
is the concatenation of the obvious projection of $\hat \Omega \frg$ onto $\Omega \frg$ and the embedding of the latter into $P_0 \frg$. The only other non-trivial product is the binary one, which is mostly the obvious commutator with a twist by the symplectic form on $\Omega \frg$:
\begin{equation}
 \begin{aligned}
  \mu_2&:P_0\frg\wedge P_0\frg\rightarrow P_0\frg~,\\
  &\hspace{0.5cm}\mu_2(\gamma_1,\gamma_2)=[\gamma_1,\gamma_2]~,\\
  \mu_2&:P_0\frg\otimes(\Omega\frg\oplus\FR)\rightarrow \Omega\frg\oplus\FR~,~~~\\
  &\hspace{0.5cm}\mu_2\big(\gamma,(\lambda,r)\big)=\left([\gamma,\lambda]\; ,\; -2\int_0^1 \dd\tau \left(\gamma(\tau),\dder{\tau}\lambda(\tau)\right)\right)~.
 \end{aligned}
\end{equation}
\end{subequations}
The quasi-isomorphism is not too hard to find~\cite{Baez:2005sn}. In particular, it is easy to see that the cohomology of $\astring_\Omega(\frg)$ is $\astring_{\rm sk}(\frg)$: two based paths with the same endpoint differ by a based loop. Moreover, $\astring_\Omega(\frg)$ is a crossed module of Lie algebras which can be readily integrated to a crossed module of Lie groups. This gives a useful 2-group model of the string group~\cite{Baez:2005sn}.

We have now two extreme examples of models of the string Lie 2-algebra: one being skeletal and thus minimal, the other being strict and thus simpler in its algebraic structure. According to our above stated principle, we expect that our higher gauge theories based on string structures can be formulated using either model. 

The last point implies that we also need a formulation of string structures for $\astring_\Omega(\frg)$. This was developed in~\cite{Saemann:2017rjm} and we have the following data. The gauge potential decomposes into
\begin{subequations}\label{eq:string_structure_loop}
\begin{equation}\label{eq:sds_fc_loop}
\begin{aligned}
 A&\in \Omega^1(\FR^4)\otimes P_0\frg~,\\
 B&\in \Omega^2(\FR^4)\otimes (\Omega\frg\oplus \FR)
\end{aligned}
\end{equation}
and the corresponding curvature components are 
\begin{equation}\label{eq:twisted_H_loop}
\begin{aligned}
 \CF &\coloneqq \dd A+\tfrac12\mu_2(A,A)+\mu_1(B)~,\\
 H&\coloneqq \dd B+\mu_2(A,B)-\kappa(A,\CF)~,
\end{aligned}
\end{equation}
where the additional map $\kappa$ is given by 
\begin{equation}
\begin{aligned}
 \kappa:P_0\frg\times P_0\frg&\rightarrow \Omega\frg\oplus \FR~,\\
 \kappa(\gamma_1,\gamma_2)&\coloneqq \left(\chi([\gamma_1,\gamma_2])\;,\;2\int_0^1\dd \tau (\dot \gamma_1,\gamma_2)\right)~.
\end{aligned}
\end{equation}
Here, $\chi(\gamma)=(\gamma-f(\tau)\dpar \gamma,0)$ for some choice of smooth function $f:[0,1]\rightarrow \FR$ with $f(0)=0$ and $f(1)=1$. The gauge transformations of the components of the gauge potential and the curvature then read as 
\begin{equation}
 \begin{aligned}
  \delta A&=\dd \alpha+\mu_2(A,\alpha)-\mu_1(\Lambda)~,\\
  \delta B&=\dd \Lambda+\mu_2(A,\Lambda)-\mu_2(\alpha,B)+\kappa(\alpha,\CF)~,\\
  \delta \CF&=\mu_2(\CF,\alpha)+\mu_1(\kappa(\alpha,\CF))~,\\
  \delta H&=0~.
 \end{aligned}
\end{equation} 
\end{subequations}
As shown in~\cite{Saemann:2017rjm}, gauge equivalence classes for both models of the string Lie 2-algebra are mapped into each other under the quasi-isomorphism linking both. Note that this form of string structure can certainly be extended to include contributions from the tangent bundle as in~\eqref{eq:string_structure_skeletal}. Here, however, we are merely interested in flat space, and we allow ourselves to simplify the discussion a bit.

\subsection{Dynamical principles}

Before continuing with constructing a 6d superconformal field theory based on string structures, let us complete the discussion of self-dual strings. Given the explicit form of string structures~\eqref{eq:string_structure_skeletal} and~\eqref{eq:string_structure_loop}, it is not too hard to come up with a generalization of the Abelian equation $H=\dd B=*\dd \phi$. 

We start from a string structure on $M=\FR^4$ for $\frg=\aspin(3)\cong \asu(2)$ as motivated above, endowed with positive definite metric originating from the Cartan--Killing form.

In the skeletal case, we add a Higgs field $\varphi\in \Omega^0(\FR^4)$ and drop the contribution from the tangent bundle (since we are on flat space\footnote{For the complete topological picture after compactification to $S^4$, one certainly needs to include this contribution.}). We impose the equation
\begin{subequations}\label{eq:sds_skeletal}
\begin{equation}
 H\coloneqq \dd B-(A,\dd A)-\tfrac{1}{3}(A,[A,A])=*\dd \varphi~,
\end{equation}
which implies 
\begin{equation}
 *\dd H=-*(\CF,\CF)=\square \varphi~,
\end{equation}
together with the Bianchi identity. The Higgs field $\varphi$ is thus determined by the second Chern character. It remains to give a dynamical equation for the curvature component $\CF$. If we assume that, just as in the case of the non-Abelian monopole, the Higgs field should suffice to recover the self-dual string up to gauge equivalence, we are led to imposing the instanton equation
\begin{equation}
 F=*F
\end{equation}
\end{subequations}
since $\varphi$ knows about the second Chern character. These equations are indeed gauge covariant if we postulate $\delta \varphi=0$.

In the loop case, we add a Higgs field $\varphi\in \Omega^0(\FR^4)\otimes (\Omega \frg\oplus \FR)$ to the string structure~\eqref{eq:string_structure_loop} for $\frg=\asu(2)$ on $\FR^4$. The equations here read as
\begin{equation}\label{eq:sds_loop}
 \begin{aligned}
  \CF&=*\CF~,\\
  H&=*\nabla \varphi=*(\dd \varphi+\mu_2(A,\varphi))~,\\
  \mu_1(\varphi)&=0~.
 \end{aligned}
\end{equation}
Again, these equations are gauge covariant.

We note that solutions to~\eqref{eq:sds_skeletal} are mapped to solutions to~\eqref{eq:sds_loop} and vice versa by the quasi-isomorphism connecting the underlying models of the string Lie 2-algebra~\cite{Saemann:2017rjm}. Altogether, we thus succeeded in formulating a dynamical principle for categorified monopoles, which respects the quasi-isomorphisms between the models of the string Lie 2-algebra. 

We close with a short discussion of the properties of our non-Abelian version of self-dual strings. On the positive side, we have the mathematical consistency of our equations following compatibility with quasi-isomor- \linebreak phisms. This implies that our equations really have non-Abelian versions of gerbes as underlying global geometric structure. Interesting topological structures can indeed be found, and the equations can be compactified on $S^4$~\cite{Saemann:2017rjm}. 

From a physics perspective, we note that our equations nicely reduce to the Bogomolny monopole equation on $\FR^3$~\cite{Saemann:2017rjm}, as one would expect from the general M-theory picture. We can also use the usual Bogomolny trick to obtain a Bogomolny bound. Moreover, the involvement of the string Lie 2-algebra, which is motivated from many perspectives in string theory is very encouraging. We shall also see that the skeletal version of our self-dual string equation appears indeed as the BPS equation of a six-dimensional superconformal field theory we shall consider later.

On the negative side, the skeletal form of the equations does not look particularly exciting: we just have a scalar field $\varphi$ in the background of an instanton and an Abelian self-dual string.

We stress, however, that, at least to our knowledge, the above yields the first non-trivial dynamical principle for a physically relevant non-trivial non-Abelian gerbe after compactification to $S^4$.

\section{Superconformal field theories in 6d}

\subsection{Cyclic structure on the string Lie 2-algebra}

We now want to use what we learned from the non-Abelian self-dual string to construct a six-dimensional superconformal field theory. We would like to be ambitious and try to write down an action principle for this theory. To this end, we first need to construct the analogue of an inner product for the string Lie 2-algebra.

An inner product on an $L_\infty$-algebra $\sL$ is a graded symmetric non-degenerate bilinear form
\begin{equation}
 (-,-):\sL\times \sL\rightarrow \FR
\end{equation}
which satisfies the obvious generalization of the usual compatibility condition for a metric Lie algebra,
\begin{equation}
 (\ell_1,\mu_i(\ell_2,\dots,\ell_{1+i}))=\pm(\ell_2,\mu_i(\ell_3,\dots,\ell_{1+i},\ell_1))
\end{equation}
for all $\ell_1,\dots,\ell_{1+i}\in \sL$, where the sign arises from graded permutation. Such inner products are known as {\em cyclic structures} and the analogue of a metric Lie algebra is often called a {\em cyclic $L_\infty$-algebra}. It can be shown that cyclic structures arise from symplectic forms of homogeneous degree on the graded vector space underlying $\sL$ shifted by $-1$.\footnote{For full details, see e.g.~\cite{Jurco:2018sby}.} 

Above, we mentioned that a string structure is a lift of the structure group of a principal spin bundle, involving a trivialization of the Chern--Simons 2-gerbe. The string Lie 2-algebra $\astring_{\rm sk}(\frg)$ is therefore better extended to the following Lie 3-algebra concentrated in degrees~$-2,-1,0$:
\begin{equation}\label{eq:string_Lie_3}
 \widehat{\astring}_{\rm sk}(\frg)=(~\FR \xrightarrow{~\id~} \FR \xrightarrow{~0~}\frg~)~,
\end{equation}
where we indicated the differential $\mu_1$. Note that $\widehat{\astring}(\frg)$ is, in fact, quasi-isomorphic to $\frg$ and it underlies the lifting construction involving the Chern--Simons\linebreak 2-gerbe~\cite{Sati:2009ic}.\footnote{This is another example of the previously mentioned necessity of choosing a model that is ``large enough.''}

Clearly, $\widehat{\astring}_{\rm sk}(\frg)$ does not allow for a homogeneous symplectic form, even after a shift by $1$. We therefore need to extend this Lie 3-algebra further, and the simplest possibility is the extension to the cotangent bundle of this graded vector space, which we then further extend to a Lie 4-algebra:
\begin{subequations}\label{eq:string_cyclic_Lie_4}
\begin{equation}\label{eq:string_cyclic_Lie_4_gv}
 \widehat{\astring}^\omega_{\rm sk}(\frg)=\xymatrixcolsep{2.4pc}
\xymatrixrowsep{1pc}
\myxymatrix{
\frg^*_c\ar@{->}[r]^{\mu_1=\id} & \frg^*_b\ar@{}[d]|{\oplus} & \FR^*_s\ar@{}[d]|{\oplus} \ar@{->}[r]^{\mu_1=\id} & \FR_p^*\ar@{}[d]|{\oplus}& \\
 & \FR_q \ar@{->}[r]^{\mu_1=\id} & \FR_r & \frg_a &}
\end{equation}
where the subscripts $a,b,c,p,q,r,s,$ are merely there to help distinguish the different summands below. Note that $\widehat{\astring}^\omega_{\rm sk}(\frg)$ is still quasi-isomorphic to $\frg$, but we now have an obvious pairing between elements of degree~$-2$ and~$0$ and between two elements of degree~$-1$. The higher products of $\astring_{\rm sk}(\frg)$ extend in a straightforward manner to $\widehat{\astring}^\omega_{\rm sk}(\frg)$, with a unique simplest extension. This extension is essentially the dual of the one used in extending the classical BRST operator to the BV action satisfying the classical master equation.

Explicitly, we have the differentials indicated in~\eqref{eq:string_cyclic_Lie_4_gv} together with the additional products 
\begin{equation}
\begin{aligned}
 &\mu_2:\frg^{\wedge 2}_a\rightarrow \frg_a~,~~~&&\mu_2(a_1,a_2)=[a_1,a_2]~,\\
 &\mu_2:\frg_a\wedge\frg^*_b\rightarrow \frg^*_b~,~~~&&\mu_2(a_1,b)=b\big([-,a_1]\big)~,\\
 &\mu_2:\frg_a\wedge\frg_c^*\rightarrow \frg_c^*~,~~~&&\mu_2(a_1,c)=c\big([-,a_1]\big)~,\\
 &\mu_3:\frg^{\wedge 3}_a\rightarrow \FR_r~,~~~&&\mu_3(a_1,a_2,a_3)=(a_1,[a_2,a_3])~,\\
 &\mu_3:\frg_a\wedge \frg_a \wedge \FR_s\rightarrow \frg^*_b~,~~~&&\mu_3(a_1,a_2,s):=s\big(\,(-,[a_1,a_2])\,\big)~,
\end{aligned}
\end{equation}
as well as the additional maps
\begin{equation}
\begin{aligned}
 &\nu_2:\frg_a\times_a \frg \rightarrow \FR_r~,~~~&&\nu_2(a_1,a_2)=(a_1,a_2)~,\\
 &\nu_2:\frg_a\times \FR^*_s\rightarrow \frg^*_b~,~~~&&\nu_2(a_1,s)=2s\big(\,(-,a_1)\,\big)~,\\
 &\nu_2:\frg_a\times \frg^*_b\rightarrow \frg^*_c~,~~~&&\nu_2(a_1,b)=b\big(\,(a_1,-)\,)
\end{aligned}
\end{equation}
for $a_{1,2,3}\in \frg_a$, $b\in \frg^*_b$, $c\in \frg^*_c$, and $s\in \FR^*_s$.
\end{subequations}

A preferred concrete choice for $\frg$ would certainly be $\frg=\asu(2)$, which we made also in the case of the non-Abelian self-dual string and which led us to the Lie 2-algebra $\astring(3)$. To get closer to the M2-brane models, we could also choose the gauge Lie algebra of the BLG-model~\cite{Bagger:2007jr,Gustavsson:2007vu}, $\frg=\asu(2)\oplus \asu(2)$, endowed with a metric of split form. That is, the metric is given by the Cartan--Killing form on the $\asu(2)$ summands with positive sign on the first and negative sign on the second summand.

\subsection{Field content and action}

One could now proceed in the manner that M2-brane models have been constructed: by postulating non-\linebreak Abelian generalizations of the Abelian supersymmetry transformations of six-dimensional superconformal field theories and deriving equations of motions as closure conditions of these transformations. Fortunately, this has already been done for us in~\cite{Samtleben:2011fj}. Here, non-Abelian extensions were inspired by tensor hierarchies of supergravity, which led to an algebraic structure governed by a set of structure constants satisfying a number of axioms. Interestingly, our Lie 4-algebra $\widehat{\astring}^\omega_{\rm sk}(\frg)$ satisfies these axioms~\cite{Saemann:2017rjm,Saemann:2017zpd}. This is very encouraging, as it gives us a link between supergravity and our action and a similar link was also present in the case of the Chern--Simons matter actions of M2-brane models.

The field content of the theory of~\cite{Samtleben:2011fj} contains an $\CN=(1,0)$ vector supermultiplet in six dimensions, and it is therefore fundamentally different from the $(2,0)$-theory. As stated above, however, our preliminary aim is merely to construct a 6d superconformal field theory which shares features with the $(2,0)$-theory. The field content of the gauge sector is as follows:
\begin{center}
\begin{proptabular}[1cm]{c|c|c}{Gauge field content}%
Field & Multiplet & Details\\
$A=A_a+A_p$ & vector  & $\in \Omega^1(\FR^{1,5})\otimes (\frg_a \oplus \FR^*_p)$\\
$\lambda^i=\lambda^i_a+\lambda^i_p$ & vector & 2 $(\frg_a \oplus \FR^*_p)$-vald.~MW spinors\\
$Y^{ij}=Y^{ij}_a+Y^{ij}_p$ & vector & 3 $(\frg_a \oplus \FR^*_p)$-vald.~aux.~scalars\\
$B=B_s+B_r$ & tensor & $\in\Omega^2(\FR^{1,5})\otimes (\FR^*_s\oplus \FR_r)$\\
$\chi^i=\chi^i_s+\chi^i_r$ & tensor & 2 $\FR^*_s\oplus \FR_r$-vald.~MW spinors\\
$\phi=\phi_s+\phi_r$ & tensor & $\FR^*_s\oplus \FR_r$-vald.~scalar field\\
$C=C_b+C_q$ & none & $\in\Omega^3(\FR^{1,5})\otimes (\frg^*_b\oplus \FR_q)$\\
$D$ & none & non-dynamical $\frg^*_e$-vald.~4-form\\
$v$ & none & PST auxiliary 1-form
\end{proptabular}
\end{center}
where $i\in\{1,2\}$. This gauge field content is complemented by matter fields in an $\CN=(1,0)$-hypermultiplet. To simplify, we restrict to $\frg=\asu(N)$, where we have
\begin{center}
\begin{proptabular}[1cm]{c|c|c}{Matter field content}%
Field & Multiplet & Details\\
$q^{ia}$ & hyper & $\FR^{2\times 2N^2}$ scalar fields\\
$\psi^a$ & hyper & $2N^2$ symplectic Majorana spinors
\end{proptabular}
\end{center}
with $a=1,\dots,2N^2$. The curvatures and covariant derivatives are given by 
\begin{equation}
 \begin{aligned}
 \CF&:=\dd A+\tfrac12 \mu_2(A,A)+\mu_1(B)~,\\
 \CH&:=\dd B-\nu_2(A,\dd A)-\tfrac13\nu_2(A,\mu_2(A,A))+\mu_1(C)~,\\
 \CG&:=\dd C+\mu_2(A,C)+\nu_2(\CF,B)+\mu_1(D)~,\\
 \CI&:=\dd D+\nu_2(\CF,C)+\dots~,\\
 \nabla \phi&:=\dd \phi+\mu_2(A,\phi)~,\\
 \nabla Y^{ij}&:=\dd Y^{ij}+\mu_2(Y^{ij})~,\\
 \nabla q^i&:=\dd q^i+ A\acton q^i~,
\end{aligned}
\end{equation}
where $\dots$ stands for additional terms irrelevant for our discussion, and $A\acton q^i$ denotes the representation of $\asu(N)$ on $\FR^{2N^2}$ obtained by an embedding $\au(n)\hookrightarrow\mathfrak{sp}(n)$, see~\cite{Saemann:2017zpd} for details. The covariant derivatives on the spinors in the vector, tensor, and hypermultiplets follow trivially.

The supersymmetry transformations are parameterized by a doublet of chiral spinors $\eps^i$ and act according to
\begin{equation}
 \begin{aligned}
  \delta A&=-\bar \eps \gamma_{(1)} \lambda~,\\
  \delta \lambda^i&=\tfrac{1}{8}\gamma^{\mu\nu}\CF_{\mu\nu}\eps^i~,\\
  \delta Y^{ij}&=-\bar \eps^{(i}\gamma^\mu \nabla_\mu \lambda^{j)}+2\mu_1(\bar\eps^{(i}\chi^{j)})~,\\
  \delta \phi&=\bar \eps \chi~,\\
  \delta \chi^i&=\tfrac{1}{48}\gamma^{\mu\nu\rho} \CH_{\mu\nu\rho}\eps^i+\tfrac{1}{4}\slasha{\dd}\phi\eps^i+\tfrac12 \nu_2(\gamma^\mu\lambda^i,\bar \eps \gamma_\mu \lambda)~,\\
  \Delta B&=-\bar \eps \gamma_{(2)}\chi~,\\
  \Delta C&=\nu_2(\bar \eps \gamma_{(3)}\lambda,\phi)~,\\
  \delta q^{ia}&=\epsb^i\psi^a~,\\
  \delta \psi^a&=\tfrac12 \slasha{\nabla}q^{ia}\eps_i~.
 \end{aligned}
\end{equation}
Here, $\mu=0,\dots,5$ are the Lorentz indices on $\FR^{1,5}$ and $\Delta$ denotes the following simplifying variation, cf.~\cite{Samtleben:2011fj}:
\begin{equation}
\begin{aligned}
 \Delta B\coloneqq\delta B-\nu_2(\delta A,A)~,\\
 \Delta C\coloneqq\delta C+\nu_2(\delta A,B)~,\\
 \Delta D\coloneqq\delta D-\nu_2(\delta A,C)~.
\end{aligned}
\end{equation}

The action is now composed of three parts:
\begin{equation}\label{eq:action}
 S=\int_{\FR^{1,6}} \CL_{\rm gauge}+\CL_{\rm matter} + \CL_{\rm PST}~,
\end{equation}
where the first Lagrangian contains the gauge fields leading to the equations of motions except for the duality conditions on the curvature forms, the second one contains the terms for the matter fields minimally coupled to the gauge sector and the last term is a PST-type action, which leads to the appropriate self-duality equations of the curvature forms. The first two terms are specializations of general actions found in~\cite{Samtleben:2011fj} and~\cite{Samtleben:2012fb}, respectively. The last term was constructed in~\cite{Saemann:2017zpd}. While a PST action for the bosonic part of $\CL_{\rm gauge}$ was given before in~\cite{Bandos:2013jva}, the supersymmetric extension was announced but not completed. From our computations, it seems that a supersymmetric completion is not possible unless one restricts the very general gauge structure of~\cite{Samtleben:2011fj} to our extended string Lie-4 algebra $\widehat{\astring}^\omega_{\rm sk}(\frg)$.

Explicitly, we have the following terms
\begin{equation}
 \begin{aligned}
  &\CL_{\rm gauge} = - \langle\dd \phi,*\dd \phi\rangle -4\vol \langle\bar\chi,\slasha{\dd}\chi\rangle-\tfrac12\langle\CH,*\CH\rangle+\\
  &\hspace{0.3cm}+\langle\CH,\nu_2(\bar\lambda,*\gamma_{(3)}\lambda)\rangle+\langle\mu_1(C),\CH\rangle +\big\langle B,\nu_2(\CF,\CF)\big\rangle-\\
  &\hspace{0.3cm}-2\big\langle\phi\,,\,\nu_2(\CF,*\CF)-2\vol~\nu_2(Y_{ij},Y^{ij})+4\vol~\nu_2(\bar\lambda,\slasha{\nabla} \lambda)\big\rangle\\
&\phantom{{}={}}+8\big\langle \nu_2(\bar\lambda,\CF),*\gamma_{(2)}\chi\big\rangle -16\vol\big\langle \nu_2(Y_{ij},\bar\lambda^i),\chi^j\big\rangle~,\\
&\CL_{\rm hyper} = -\langlec \nabla q,*\nabla q\ranglec + 2\vol\langlec\bar\psi,\slasha{\nabla}\psi\ranglec +\\
&\hspace{0.3cm}+8\vol\langlec\bar\psi, \lambda_i\acton q^i\ranglec + 2\vol\,\langlec q^i,Y_{ij}\acton q^j\ranglec~,\\
&\CL_{\rm PST} = \tfrac12\big\langle\iota_{V}\CH^+,\CH^+\big\rangle\wedge v+\langle \Phi(\iota_{V} * \CG^+), *~\iota_{V}*\CG^+\rangle~.
\end{aligned}
\end{equation}
Here, the vector field $V$ is defined by $\iota_V v=1$ and $\iota_V *v=0$. Moreover, we abbreviated notation by using the supersym\-metrically covariant components of the curvatures
\begin{equation}
\begin{aligned}
 \CH^+&:=*\CH-\CH-\nu_2(\bar \lambda,\gamma_{(3)}\lambda)~,\\
 \CG^+&:=\CG-\nu_2(*\CF,\phi)+2\nu_2(\bar \lambda,*\gamma_{(2)}\chi)~,\\
 \CI^+&:=\CI+\mu_2(\nu_2(\bar\lambda,\phi),\gamma^\mu\lambda)\vol_\mu+2\langlec-\acton q,*\nabla q\ranglec-\\
 &\hspace{0.6cm}-2\langlec\bar\psi,-\acton\gamma^\mu\psi\ranglec\vol_\mu~,
\end{aligned}
\end{equation}
where $\vol_\mu=\iota_{\der{x^\mu}}\vol$ denotes the evident contraction of the volume form on $\FR^{1,5}$ by $\der{x^\mu}$. It is these curvature components that are put to zero by the equations of motion.

\subsection{Properties of the action}

Our action~\eqref{eq:action} is a six-dimensional superconformal field theory which contains a 2-form potential whose equations of motion fix the anti-self-dual part of the 3-form to $\CH^+=0$ via a generalization of the PST mechanism. 

Its field content is that of the $(2,0)$-theory plus an $\CN=(1,0)$ vector multiplet. The latter prohibits $\CN=(2,0)$ supersymmetry. Note that reduced manifest supersymmetry is also a generic feature of M2-brane models. These are $\CN=6$ supersymmetric and full $\CN=8$ supersymmetry is only recovered after introducing monopole operators. There is a reasonably clear higher analogue of the latter in the form of ``self-dual string operators,'' however, it is not clear if they can ameliorate the situation. Another parallel to M2-brane models is that the interactions arise from coupling an essentially free theory to an additional gauge potential. 

The action is based on string structures, which implies that there is the clean mathematical language of higher principal bundles behind it, and at least its gauge part can be put on any suitable manifold, as one would expect. It also means that the gauge structure is reasonably natural. In particular, our action exists for $\frg$ any of the Lie algebras of types A, D, or E. A restriction to those, however, is not seen in the classical case.

The action is clearly non-trivial, interacting and it contains the Abelian tensor multiplet as a special case. The action also contains higher Chern--Simons terms, in particular the terms
\begin{equation}
 \langle\mu_1(C),\CH\rangle +\big\langle B,\nu_2(\CF,\CF)\big\rangle
\end{equation}
in $\CL_{\rm gauge}$ are topological. The coupling constant, which we suppressed here, is therefore quantized, which provides an explanation for the lack of continuous coupling constants or, equivalently, continuous deformations of the action for an Abelian self-dual 3-form to an interacting one.

We suppressed the supersymmetry transformations in the previous section, but if we consider the transformations of the fermions,
\begin{equation}
 \begin{aligned}
    \delta \chi^i&=\tfrac{1}{48}\gamma^{\mu\nu\rho} \CH_{\mu\nu\rho}\eps^i+\tfrac{1}{4}\slasha{\dd}\phi\eps^i+\tfrac12 \nu_2(\gamma^\mu\lambda^i,\bar \eps \gamma_\mu \lambda)~,\\
    \delta \lambda^i&=\tfrac{1}{8}\gamma^{\mu\nu}\CF_{\mu\nu}\eps^i~,
 \end{aligned}
\end{equation}
we see that after a dimensional reduction from $\FR^{1,5}$ to $\FR^4$, the BPS equations are indeed the equations of the skeletal self-dual string~\eqref{eq:sds_skeletal}. Here, the scalar field $\varphi$ in the self-dual string equation is obtained by the component of the 2-form field $B$ along the dimensionally reduced directions.

A crucial test of any theory that claims to be close to the $(2,0)$-theory is a consistent reduction to five-dimen\-sional maximally supersymmetric Yang--Mills theory. While it is not clear to us how this reduction could be performed in our action~\eqref{eq:action}, there is a straight-forward way of obtaining four-dimensional super Yang--Mills theory including the $\theta$-term, even if its physical motivation is rather unclear\footnote{It is slightly similar to the reduction of M2-brane models to super Yang--Mills theories for D2-branes in that scalar fields acquire an expectation value.}. We assume that the fields $\phi_s$ and $B_s$ acquire the following expectation values:
\begin{equation}
 \begin{aligned}
 \langle \phi_s \rangle&=-\frac{1}{32\pi^2} \frac{1}{R_{10}^2}=-\frac{1}{32\pi^2}\frac{\tau_2}{R_9 R_{10}}~,\\
 \langle B_s \rangle&=\frac{1}{16\pi^2}\frac{\tau_1}{R_9 R_{10}} {\rm vol}(T^2)~,
 \end{aligned}
\end{equation}
where
\begin{equation}
 \tau=\tau_1+\di \tau_2=\frac{\theta}{2\pi}+\frac{\di}{g^2_{\rm YM}}
\end{equation}
is the modular parameter of the torus used in the compactification from $\FR^{1,5}$ to $\FR^{1,3}$. After integrating over the radial directions, assuming constancy of all fields in these, and performing a strong coupling expansion, we recover the action of four-dimensional super Yang--Mills theory.

T-duality in string theory suggests that there may be a remnant of this symmetry within M-theory linking M5-branes to M2-branes. It is therefore desirable to see if our action can be reduced to a Chern--Simons matter theory in three dimensions. Let us compactify $\FR^{1,5}$ to $\FR^{1,2}\times M$, where $M$ is some compact 3-manifold, e.g.~$M=T^3$. We assume that the gauge potential 1-form $A$ vanishes on $M$ and that the 2-form $B$ vanishes on $\FR^{1,2}$, that is, $B$ is part of the connective structure of a gerbe with Dixmier--Douady class $k$ over $M$. The integral in the action then reduces according to 
\begin{equation}
\begin{aligned}
 \int_{\FR^{1,2}\times M} \CH\wedge * \CH&= \int_M \dd B~~ \int_{\FR^{1,2}} {\rm cs}(A)\\
 &= k \int_{\FR^{1,2}} {\rm cs}(A)~,
\end{aligned}
\end{equation}
and we recover in total a supersymmetric Chern--Simons matter action together with a quantized Chern--Simons level.

Besides these very encouraging properties of our action, let us mention some of the less desirable features. First of all, the theory relies on an $\CN=(1,0)$ vector multiplet, which renders an enhancement of supersymmetry to $\CN=(2,0)$ impossible. Second, the action as given could be regarded as suffering from unitarity problems because the term $\langle \phi_s,\nu_2(\CF,*\CF)\rangle$ is clearly not bounded from below. Similarly, the PST formalism which we used relies on the fact that $\phi_s>0$. Both issues can be remedied by restricting $\phi_s>0$. The physical interpretation of the value of $\phi_s$ is the distance of the two M5-branes whose relative motion might be described by our action~\eqref{eq:action} and the model should not care about the sign. Nevertheless, a better and cleaner interpretation of both related issues should be sought. Clearly, the $(2,0)$-theory sits at the point $\phi_s=0$, which is highly problematic in our action.

There are a number of further issues with our theory and a detailed criticism can be found in~\cite{Saemann:2017zpd}. 

There is also a fundamental open problem with our setup. Separating a stack of $N=N_1+N_2$ D-branes into two stacks of $N_1$ and $N_2$ D-branes in string theory is (in the simplest case) reflected by a branching of the Lie algebra $\au(N_1+N_2)\rightarrow \au(N_1)\oplus \au(N_2)$. It is not clear whether a similar splitting that matches all expectations exists for the string Lie 2-algebra or its simple generalizations. It may be that this Lie 2-algebra is only suitable for pairs of M5-branes, where the branching issue is absent\footnote{Our action may be seen as describing the relative motion of two M5-branes, with the center-of-mass motion to be added as an action of a free tensor multiplet.}.

\section{Conclusions}

We explored the possibility of formulating a classical action for the (2,0)-theory as a higher gauge theory. The latter framework is strongly suggested by a number of physical and mathematical arguments. The mathematics behind higher gauge theory is fully developed and there are mathematically well-defined higher analogues of gauge groups, principal bundles and connections. 

Various arguments suggest that a classical description of the (2,0)-theory may not exist, and therefore we merely aimed for a six-dimensional superconformal field theory which shares as many properties as possible with the (2,0)-theory.

We only discussed the local picture and considered $L_\infty$-algebras, which are higher versions of Lie algebras. We observed that the usual description of higher gauge theory is fine for higher analogues of Chern--Simons theories in which the curvatures vanish. In a higher gauge theory with generically non-vanishing curvatures, however, the so-called fake curvature condition is problematic in that it renders all kinematical configurations gauge equivalent to purely Abelian ones, at least over contractible spaces. This can be fully remedied by a mathematically and physically well-motivated modification of the expressions of higher curvature forms. We gave an explicit example of this modification in the form of connections and curvatures on string structures, which are essentially higher generalizations of spin bundles. 

In order to determine a suitable and interesting higher gauge group for the classical action of our theory, we turned our attention to self-dual strings. It is known that these categorified monopoles are the BPS states in the $(2,0)$-theory. While Abelian self-dual strings have been known for a long time, a nontrivial non-Abelian extension which is mathematically consistent with the framework of higher gauge theory was only developed in~\cite{Saemann:2017rjm}. The latter formulation was motivated by analogies of the relations between Abelian and non-Abelian monopoles which led to connections on string structures. In fact, 2-group models of the string group as higher gauge group are strongly supported from many perspectives within string theory.

To be able to write down an action functional, we had to introduce the notion of an inner product on the higher Lie algebra underlying string structures. Such inner products arise from symplectic forms, and we had to extend the graded vector space of the higher Lie algebra to a symplectic one. This extension is readily performed and it is dual to the extension of the BRST complex to the BV complex. 

We stress again that the resulting algebraic structure is well-motivated and the (higher) infinitesimal symmetries it describes have finite analogues. Moreover, higher principal bundles with these finite analogues as higher structure groups are mathematically well defined.

Our gauge algebraic structure is not only interesting for its mathematical consistency but also because it is a special case of a gauge structure that has been derived ``by hand'' in~\cite{Samtleben:2011fj} in the construction of six-dimensional superconformal field theories from tensor hierarchies. We can thus take this general action and specialize to string structures. This solves the question of which gauge structure to choose in the model of~\cite{Samtleben:2011fj}: string structures are (at least) particularly interesting candidates.

The resulting action now shares many properties with the $(2,0)$-theory, but also comes with some features which contradict $(2,0)$-supersymmetry. Among the former is certainly the appearance of our non-Abelian self-dual string as a BPS solution as well as the possible dimensional reductions to four-dimensional super Yang--Mills theory and three-dimensional Chern--Simons matter theories. Among the problematic features is the strong reliance on an $\CN=(1,0)$ vector supermultiplet, which seems to make a symmetry enhancement to $\CN=(2,0)$ impossible.

Altogether, we can claim partial success: there is at least one superconformal six-dimensional field theory which is truly interacting. Whether this theory can be consistently quantized depends on whether one can consistently restrict the scalar field $\phi_s$ to the range $\phi_s\in \FR^+$.

Let us highlight a further issue which is rather positive: our action~\eqref{eq:action} is {\em not} compatible with quasi-isomor\-phisms. That is, the form of this action needs to be modified significantly to arrive at the equivalent model for the Lie 2-algebra model $\widehat\astring_{\Omega}^\omega(\frg)$. This is encouraging since it underlines the mathematical incompleteness of the action~\eqref{eq:action} with which we are also not fully satisfied for physical reasons. We are currently working on this point and hope to be able to report on progress soon, see also~\cite{contrib:schmidt}.

Finally, let us stress that most of the problems we encountered with our action can be circumvented when discussing merely equations of motion. In our previous work~\cite{Saemann:2012uq,Saemann:2013pca,Jurco:2014mva,Jurco:2016qwv} we found a description of solutions to these equations in terms of holomorphic categorified principal bundles over a suitable twistor space. This essentially reduced the search for classical equations of the $(2,0)$-theory to a search for the right gauge structure (which implies a notion of holomorphic categorified principal bundle). Since string structures have been clearly identified as a primary candidate, it merely remains to construct the corresponding holomorphic categorified principal bundles and to plug them into the above twistor description. Our current work on this point should also be completed soon.
 
\bibliographystyle{prop2015}
\bibliography{allbibtex}

\end{document}